# p-Diamond, Si, GaN and InGaAs TeraFETs

Yuhui Zhang and Michael S. Shur, *Life Fellow, IEEE*

*Abstract*—p-diamond field effect transistors (FETs) featuring large effective mass, long momentum relaxation time and high carrier mobility are a superb candidate for plasmonic terahertz (THz) applications. Previous studies have shown that p-diamond plasmonic THz FETs (TeraFETs) could operate in plasmonic resonant mode at a low frequency window of 200 GHz to ~600 GHz, thus showing promising potential for beyond 5G sub-THz applications. In this work, we explore the advantages of p-diamond transistors over n-diamond, Si, GaN and InGaAs TeraFETs and estimate the minimum mobility required for the resonant plasmons. Our numerical simulation shows that the p-diamond TeraFET has a relatively low minimum resonant mobility, and thus could enable resonant detection. The diamond response characteristics can be adjusted by changing operating temperature. A decrease of temperature from 300 K to 77 K improves the detection performance of TeraFETs. At both room temperature and 77 K, the p-diamond TeraFET presents a high detection sensitivity in a large dynamic range. When the channel length is reduced to 20 nm, the p-diamond TeraFET exhibits the highest DC response among all types of TeraFETs in a large frequency window.

## I. INTRODUCTION

Recently, there has be a tremendous advance in the analysis and design of plasmonic terahertz (THz) semiconductor devices [1-6] due to the pressing demand of applications in various industrial fields. The fundamental theories of plasma-wave terahertz field-effect transistors (TeraFET) were developed in 1990s [7-9], later verified and supplemented by further experimental and analytical works [2, 3, 10, 11]. By now, THz imaging [12], sensing [13], communication [14] and optical pulse detection [6] are four mainstream applications of TeraFETs.

TeraFETs have been implemented in silicon [15], InGaAs [3], III-N [16], and graphene [17]. Recently, p-diamond has been proposed as another promising candidate for plasmonic THz applications [4]. Diamond is a wide bandgap material (~5.46 eV) and has a relatively high dielectric strength (~5 to 10 MV/cm) as well as a high thermal conductivity (about 23 W/cm·K) [4]. Those characteristics make diamond a very promising material for power and high temperature applications. Diamond can also be used for RF power device applications since it can withstand a large microwave signal

The authors are with the Department of Electrical, Computer and Systems Engineering, Rensselaer Polytechnic Institute, Troy, NY 12180 USA (e-mail: zhangy79@rpi.edu; shurm@rpi.edu).

[18]. In above applications, both delta [19, 20] and transfer doping diamond devices [21, 22] were proposed, and the operating temperature could vary from room temperature [19] to 200-250 °C [20] and 350 °C [23].

Diamond has a high optical phonon energy (165 meV [24]), which leads to the suppression of optical phonon scattering and the dominance of acoustic phonon scattering up to ~400 K [25]. This feature along with a large 2DEG/2DHG effective mass ($m_e$ = (0.28-0.36)$m_0$, $m_h$ = (0.63-2.12)$m_0$ [4, 26, 27], where $m_e$, $m_h$ and $m_0$ are the electron effective mass, hole effective mass and free electron mass, respectively) give rise to a very large momentum relaxation time $\tau$ and makes diamond easier to meet the plasmonic resonance condition $\omega_p \tau > 1$, where $\omega_p$ is the plasma frequency [4]. With a high $\tau$, the plasmonic resonance can be achieved at above ~200 GHz [4], making diamond promising for THz detection in the 240GHz and ~600 GHz windows allocated for beyond 5G THz communications [4, 13, 14, 28-30]. Diamond FETs also have a relatively low ohmic contact resistance [4], which makes diamond even more advantageous for THz electronic applications.

Compared to n-diamond, p-diamond has a larger effective mass and a comparable maximum carrier mobility (as high as ~5300 cm$^2 \cdot$V$^{-1}$s$^{-1}$ [4, 25]). Therefore, here we focus on the characteristics of p-diamond as a TeraFET material.

## II. BASIC EQUATIONS

We use one-dimensional hydrodynamic model to trace the generation and propagation of plasma waves in gated plasmonic terahertz FETs (TeraFETs) implemented in p-diamond, n-diamond Si, GaN, and InGaAs, with feature sizes (gate lengths) of 20 nm, 65 nm, and 130 nm at 300 K and 77 K. Table 1 lists the materials parameters used in the simulation. The mobilities used in our simulation are relatively large, which are close to the maximum reported values. The governing hydrodynamic model equations are [3, 11]

$$\frac{\partial n}{\partial t} + \nabla \cdot (n\boldsymbol{u}) = 0 \quad (1)$$

$$\frac{\partial \boldsymbol{u}}{\partial t} + (\boldsymbol{u} \cdot \nabla)\boldsymbol{u} + \frac{e}{m}\nabla U + \frac{\boldsymbol{u}}{\tau} - \nu\nabla^2\boldsymbol{u} = 0 \quad (2)$$

$$\frac{\partial \theta}{\partial t} + \nabla \cdot (\theta\boldsymbol{u}) - \frac{\chi}{C_v}\nabla^2\theta - \frac{m\nu}{2C_v}(\frac{\partial u_i}{\partial x_j} + \frac{\partial u_j}{\partial x_i} - \delta_{ij}\frac{\partial u_k}{\partial x_k})^2$$
$$= \frac{1}{C_v}(\frac{\partial W}{\partial t})_c + \frac{m\boldsymbol{u}^2}{C_v\tau} \quad (3)$$

Here $n$ represents the carrier density, $\boldsymbol{u}$ stands for the hydrodynamic velocity. The gate-to-channel potential $U$ is defined as $U = U_0 - U_{ch}$ where $U_0$ is the gate bias above threshold and $U_{ch}$ is the channel potential. The gradual channel approximation ($CU = en$, where $C$ is the capacitance of barrier layer and is set to 0.56 µF/cm$^2$ [3]) is used in the simulation. $\tau$ is the momentum relaxation time of carriers.

TABLE I
MATERIALS PARAMETERS USED IN THE SIMULATION

| Material | $m/m_0$ | Mobility (77 K) | Mobility (300 K) | Ref. |
|---|---|---|---|---|
| p-diamond | 0.74 | 35000 | 5300 | [4, 25, 27] |
| n-diamond | 0.36 | 50000 | 7300 | [4, 25, 31] |
| Si | 0.19 | 20000 | 1450 | [32-35] |
| GaN | 0.24 | 31691 | 2000 | [35-37] |
| InGaAs | 0.041 | 35000 | 12000 | [6, 38, 39] |

Note: $m$ represents the effective mass of carriers, $m_0$ is the mass of free electrons. The unit of mobility is cm$^2 \cdot$V$^{-1}$s$^{-1}$.

In the energy transport equation (3), $\theta = k_BT$ is the temperature in eV, $\chi$ is defined as $\chi = \kappa/n$ ($\kappa$ is the heat conductivity) and has the same units as kinematic viscosity; $C_v$ is the thermal capacitance defined as $C_v = (\partial\Sigma/\partial\theta)_n$ [6, 11], where $\Sigma = \theta*F_1(\xi)/F_0(\xi)$ is the average internal energy, $\xi = \ln(\exp(E_F/k_BT)-1)$ is the chemical potential, $E_F = k_BT_F = \pi\hbar^2 n/m$ is the Fermi energy, and $F_k(\xi)$ is the Fermi integral; $W$ is the total energy. $(\partial W/\partial t)_c$ represents the collision term of $\partial W/\partial t$, which can be expressed by $(\partial W/\partial t)_c = (\partial\Sigma/\partial t)_c - m\boldsymbol{u}^2/\tau$. Furthermore, $v$ is the viscosity of the 2DEG/2DHG, which is related to the carrier density and temperature. When the temperature $T$ is much lower than the Fermi temperature $T_F$, $\chi$ and $v$ have the following expression [3, 6, 11]:

$$v(T) = \frac{2\hbar}{\pi m}\frac{T_F^2}{T^2}\frac{1}{\ln(2T_F/T)} \quad (T \leq T_F) \quad (4)$$

$$\chi(T) = \frac{4\pi\hbar}{3m}\frac{T_F}{T}\frac{1}{\ln(2T_F/T)} \quad (T \leq T_F) \quad (5)$$

Generally, the condition $T < T_F$ holds for relatively large gate bias $U_0$ (e.g. $U_0 > 2.3$ V for p-diamond TeraFETs), at which the viscosity and heat conductivity could be large. For small $U_0$, $T > T_F$, and we use $v(T = T_F)$ and $\chi(T = T_F)$, which are constant values [3]. In this condition, the Navier-Stokes equation (2) effectively decouples from the heat transport equation (3).

The boundary conditions that we use are an open drain condition [3, 7]: $U(0, t) = U_0 + U_a(t)$ and $J(L, t) = 0$, where $U_a(t) = U_{am}\cdot\cos(\omega t)$ represents the AC voltage induced by the incoming THz radiation, $J$ is the current flux density. The above equations were established and solved in COMSOL 5.4 [40] using the finite element method. More detailed introductions of the hydrodynamic model are given in [3] and [11].

The above-mentioned hydrodynamic model [3, 11, 41, 42], is valid when the electron–electron scattering rate $1/\tau_{ee}$ is much larger than the inverse momentum relaxation time [6]. We have checked that in our simulation, the value of $\tau_{ee}$ [43] is always much smaller than the momentum relaxation time $\tau$, so that the validity of hydrodynamic equations is guaranteed. The results obtained from hydrodynamic models were compared to the experimental measurements in previous studies, and some good agreements between modeling/analytical results and experimental data were observed [3, 44-48].

The boundary conditions of an open drain represent the ideal boundary conditions for the TeraFET detectors, i.e. the load resistance $R_L$ at the drain is infinite. As shown in [49], for a finite load resistance at the drain, the response, $V_L$ becomes smaller that the open drain response $V_o$:

$$V_L = \frac{V_o}{1 + R_{ch}/R_L} \quad (6)$$

Here $R_{ch}$ is the DC TeraFET channel resistance. For above-threshold, open drain operation, $R_{ch}$ is always much smaller than $R_L$, so that $V_L \approx V_o$. For deep-subthreshold or high drain capacitance cases, $R_{ch}$ could reach tens to hundreds of megaohms [49], and $R_L$ could reduce significantly. Then the DC response would be attenuated due to the device loading effect.

In addition to the numerical modeling, we also use analytical theories to predict the DC response and compare it with modeling results. The analytical equations for DC response in TeraFETs were first established in [7] and were further modified for more detailed applications [3, 11, 45]. Here we follow the analytical equations derived in [7].

## III. RESULTS AND DISCUSSION

### A. Resonant versus non-resonant response

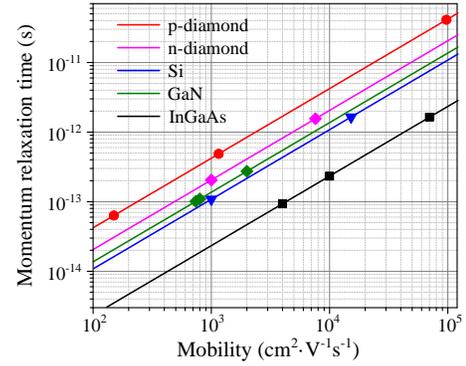

Fig. 1. Momentum relaxation time versus mobility for p-diamond, n-diamond, Si, InGaAs, GaN. Symbols represent measured values reported in: for p-diamond [4, 35], for n-diamond [4, 31], for Si [35, 50], for GaN [51, 52], for InGaAs [38, 53].

In sub-THz and THz plasmonic devices, the momentum relaxation time $\tau$ is a key parameter, as it determines the critical condition for the plasmonic resonance. Therefore, we first compare the momentum relaxation time ($\tau = \mu m/e$) versus the mobility in different material systems, as shown in Fig. 1. Clearly, the p-diamond has the highest $\tau$ compared to other materials due to a high effective mass, and the value of $\tau$ varies between ~10$^{-14}$ s to ~10$^{-11}$ s as the mobility alters from 10$^2$ cm$^2 \cdot$V$^{-1}$s$^{-1}$ to ~10$^5$ cm$^2 \cdot$V$^{-1}$s$^{-1}$. In the case of highest $\tau$, the resonance quality factor $\omega_p\tau$ could exceed unity at a very low frequency (as low as ~100 GHz or even smaller at cryogenic temperatures). Therefore, the frequency range for resonant detection in p-diamond TeraFETs could be much wider than devices made with stereotype materials, which is very promising for sub-THz communication applications.

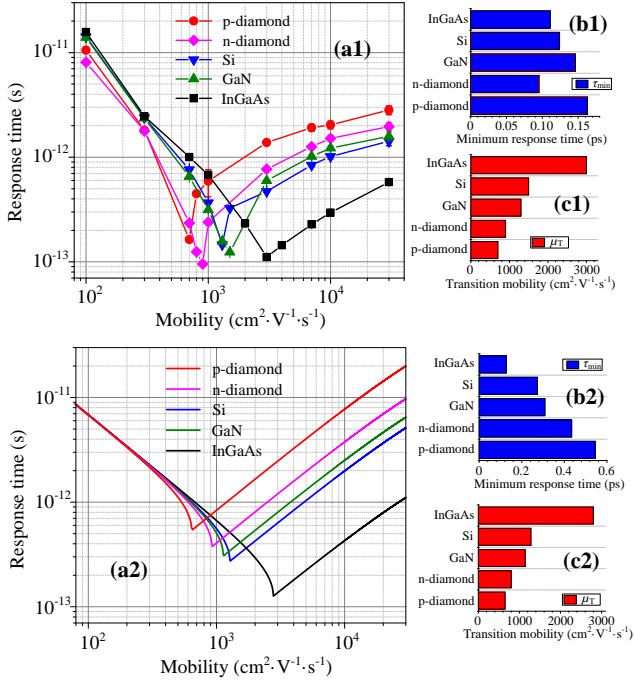

Fig. 2. Ultimate response times of TeraFETs made with 5 materials versus carrier mobility (a1), (a2), as well as their minimum response time (b1), (b2) and transition mobility (c1), (c2). (a1)-(c1) show the simulation data, and (a2)-(c2) present the analytical results. The gate voltage swing $U_0$ is set 0.1 V, the gated channel length $L$ is 130 nm, the pulse width of incoming radiation is $5\times10^{-14}$ s, under room temperate ($T$ = 300 K).

With a large effective mass, p-diamond TeraFETs could reach the plasma resonance condition $\omega_p\tau = 1$ at a relatively low frequency. It follows that at any given frequency, the critical $\tau$ (or critical mobility) required for resonant operation in p-diamond devices could also be low [4]. As reported previously, the minimal resonant mobility in TeraFETs can be evaluated by measuring the ultimate response time ($\tau_r$) of the device and finding the mobility at which the minimum $\tau_r$ is reached [6, 54-56]. Here we also adopt this method. Fig. 2 shows the ultimate response time of TeraFETs made with different materials as a function of mobility. The $\tau_r$ values are calculated by feeding an ultra-short ($5\times10^{-14}$ s in this case) square-pulse signal to the source and evaluating the resulting voltage response [6]. In this figure, we also present the analytical response time curves for comparison. The analytical theory follows [6] and [3], in which the response was fit into a oscillating decay in the form of $\sum_{(n)} A_n \exp(\sigma_n t) f_n(x)$, where $\sigma_n$ is obtained from the solution of linearized equations and is given by:

$$\sigma_n^\pm = \frac{1}{2}\left[-\left(\frac{1}{\tau}+\frac{\pi^2 v n^2}{4L^2}\right) \pm \sqrt{\left(\frac{1}{\tau}+\frac{\pi^2 v n^2}{4L^2}\right)^2 - \frac{\pi^2 s^2 n^2}{L^2}}\right] \quad (6)$$

$n = 1, 3, 5, ...$

The response time of the periodic oscillation can be obtained from the first-order approximation, i.e. $\tau_r = 1/\text{Re}(|\sigma_1^+|)$.

As shown in Fig. 2(a1), for all materials, the response time initially decreases with $\mu$, reaches a minimum, and then increases and gradually saturates. In the low mobility regime, the plasma waves are overdamped, and the voltage response is purely an exponential decay. The ultimate response time in this regime is on the order of $L/\mu U_0$ [6]. The high-mobility region corresponds to the region of resonant plasmonic operation. The minimum $\tau_r$ is reached at the transition mobility between two regimes [6, 54]. The general variation trend of $\tau_r$ shown in Fig. 2(a1) agrees with the analytical results in Fig. 2(a2). For transition mobility $\mu_T$, as presented in Fig. 2(c1), p-diamond TeraFET has the lowest $\mu_T$ among 5 material systems, and the value of $\mu_T$ for p-diamond (700 cm$^2\cdot$V$^{-1}$s$^{-1}$) is much lower than those of GaN, Si and InGaAs transistors. It is also observed that the simulation values of $\mu_T$ conform well to the analytical data shown in Fig. 2(c2). Therefore, the plasmonic resonance is easier to be induced in the p-diamond TeraFET compared to other TeraFETs. This could be a huge advantage for p-diamond over other materials in making TeraFETs, since the realization of high carrier mobility could require substantial efforts.

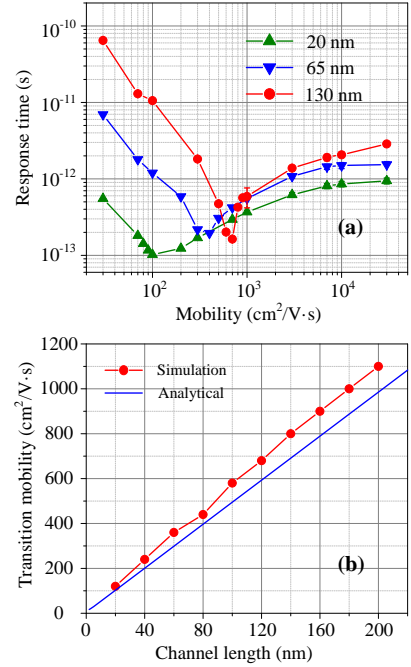

Fig. 3. Ultimate response time of the p-diamond TeraFET as a function of carrier mobility under 3 different channel lengths (a), as well as the transition mobility versus channel length for the same FET (b). The gate voltage swing $U_0$ is set 0.1 V and the pulse width of incoming radiation is $5\times10^{-14}$ s, under room temperate.

In addition, it is also interesting to investigate the variation of ultimate response time and minimum resonant mobility with the channel length $L$. The results are presented in Fig. 3. As shown in Fig. 3(a), as $L$ alters, the transition mobility and minimum $\tau_r$ reduces with decreasing $L$. This suggests that small-feature-size TeraFETs could be advantageous over those with large feature sizes. Fig. 3(b) presents the minimum mobility required for resonant plasmonic operation versus the gated channel length in p-diamond TeraFET. As seen, the critical mobility decreases almost linearly with the reduction of channel length. According to previous theoretical analyses, the transition mobility $\mu_T$ is on the order of $(eL/\pi ms)/(1-\pi v/4Ls)$, and the corresponding analytical curve is also given in Fig. 3(b). Obviously, the analytical curve exhibits a linear feature. This is because the contribution of term $(1-\pi v/4Ls)$ is negligible in this case and thus the analytical value of $\mu_T$ is very close to $Le/\pi ms$, which has a linear dependence to $L$. Moreover, the simulation values are in good agreement with the theoretical results. The conformity between simulation and analytical results enables us to predict the minimum resonant mobility with the analytical equations obtained previously.

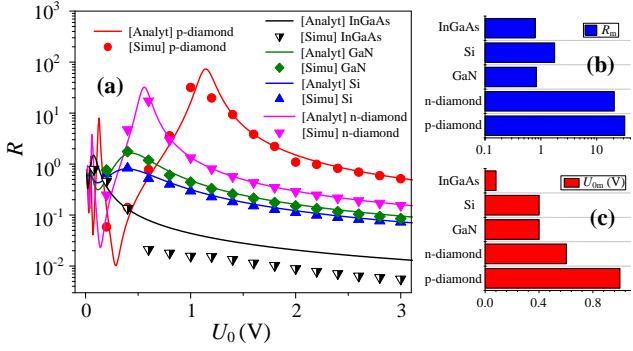

Fig. 4. Normalized response of different TeraFETs versus the gate bias $U_0$ (a), along with the amplitudes (b) and positions (c) of peak response. The frequency and amplitude of incoming small signal are set 1 THz and 5 mV, respectively. The channel length $L = 130$ nm, under room temperate.

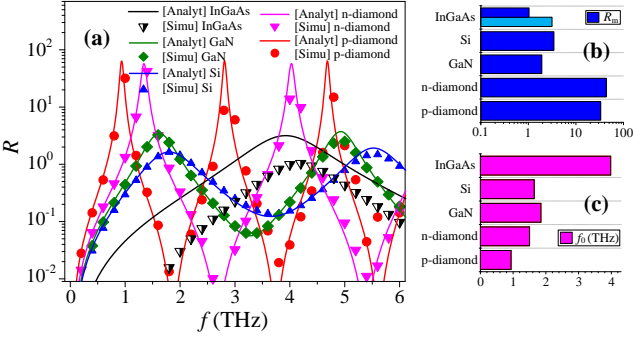

Fig. 5. Normalized response of different TeraFETs versus frequency $f$ (a), along with the amplitudes of peak response ($R_m$) (b) and the values of fundamental frequency ($f_0 = s/4L$, where $s = (eU_0/m)^{0.5}$ is the plasma velocity) (c). The cyan bar in (b) presents the $R_m$ in analytical curve. The amplitude of incoming small signal is 5 mV. The gate voltage swing $U_0$ is 1 V. The channel length $L = 130$ nm, under room temperate.

### B. General frequency-dependent and bias dependent profiles

The above discussion shows that p-diamond TeraFETs have relatively high momentum relaxation times and low minimum resonant frequencies, therefore promising for plasmonic THz applications. Now we further discuss the characteristics of p-diamond TeraFETs from the perspective of DC voltage response. Fig. 4 presents the normalized DC voltage response versus gate bias for 5 types of plasmonic FETs under $T = 300$ K, $L = 130$ nm. Both analytical curves (solid lines) and simulation values (scatters) are depicted in this figure. The analytical DC source-to-drain voltage response is proportional to the power of incoming AC small signal [7]:

$$\frac{dU}{U_0} = \frac{1}{4}\frac{U_{am}^2}{U_0^2} f(\omega) \quad (7)$$

where $U_{am}$ is the amplitude of incoming AC signal, $f(\omega)$ is a frequency dependent function given in [7], and is always positive. Besides, the normalization of $dU$ is performed as $R = dU \cdot U_0/U_{am}^2$.

As shown in Fig. 4(a), at 300 K when the viscosity is relatively low, the simulation data show very good match with their corresponding analytical curves in p-diamond, n-diamond,

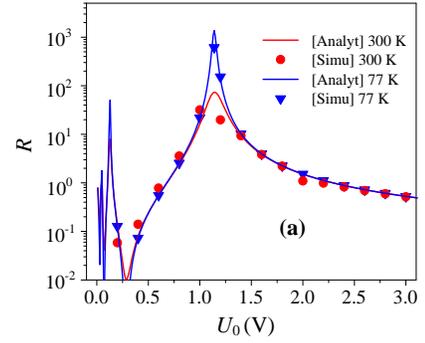

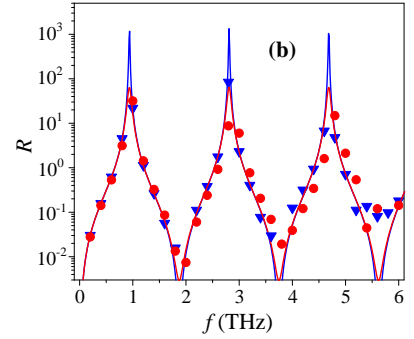

Fig. 6. Normalized response of the p-diamond TeraFET versus $U_0$ (a) and $f$ (b) under typical mobilities at 3 temperatures. The $f$ in (a) is 1 THz, and the $U_0$ in (b) is 1 V. The amplitude of incoming small signal is 5 mV. The channel length $L = 130$ nm.

Si and GaN cases. For the case of InGaAs, a significant discrepancy between simulation and analytical results is observed at $U_0 \geq 0.6$ V. Note that with a small effective mass, the viscosity of InGaAs could be remarkable even at room temperature (e.g. $v = 179.4$ cm$^2 \cdot$V$^{-1}$s$^{-1}$ at $U_0 = 0.6$ V). Therefore, this mismatch between simulation and theory is likely attributed by the high viscosity of InGaAs [3, 7, 9]. Furthermore, it can be seen from Fig. 4(b) that p-diamond TeraFET has the largest peak response, and it leads all TeraFETs in response when $U_0$ is above 0.8 V. This indicates that p-diamond detectors might be most advantages for high gate bias operation at room temperature. Besides, the position of resonant peak $U_{0r}$ can be roughly obtained by the expression of fundamental resonant frequency, as given below:

$$n \cdot f_0 = n \cdot \frac{s}{4L} = \sqrt{\frac{n^2 e U_{0r-n}}{16mL^2}} = f \Rightarrow U_{0r-n} = \frac{16mL^2 f^2}{n^2 e} \quad (8)$$

where $n = 1, 3, 5 \ldots$ is the odd harmonic ordinal. For a fixed $f$, a higher effective mass $m$ would require a larger $U_0$ to meet the criterion for plasma resonance, and this is the reason why the peak of DC response curve moves rightward with increasing effective mass. The values of fundamental $U_{0r}$ (i. e. $n = 1$) predicted by equation (8) for p-diamond, n-diamond, Si, GaN and InGaAs are 1.14 V, 0.55 V, 0.29 V, 0.37 V, 0.063 V, respectively, which in general agrees with the results of $U_{0m}$ shown in Fig. 4(c). In addition, for p- and n-diamond TeraFETs, there are some other response peaks at relatively low gate bias region, corresponding to the higher order harmonics. Since the value of $R$ varies sharply in this region, a very low $U_0$ could be undesirable for THz detection.

Apart from the $U_0$ dependence, it is also important to study the frequency dependence of the DC response in different TeraFETs. With this in mind, we simulate the normalized voltage response versus the driving frequency presented in Fig. 5. As seen, the simulation data still conforms to the analytical curves for all types of TeraFETs except InGaAs. Also, the peaks of response curves are precisely located at the fundamental frequencies (see Fig. 5(c)) and their odds harmonics, indicating all the detectors operate at resonant regime. In this case, p-diamond TeraFET has the second highest peak response, which is much large than those of GaN, Si and InGaAs TeraFETs, as shown in Fig. 5(a) and Fig. 5(b). At low frequency ($f < 1$ THz), the response of p-diamond TeraFET is the highest among all devices. This suggests that p-diamond detector might be the most suitable one for sub-THz detection, and could support the resonant plasmonic response in the 240-320 GHz band (in agreement with prediction made in [4]), which is very important in technologies related to beyond-5G communications. Beyond 1 THz, the p-diamond TeraFET also yields the highest DC response in the vicinity of its harmonic peaks. Those features allow better performance for high sensitivity detection.

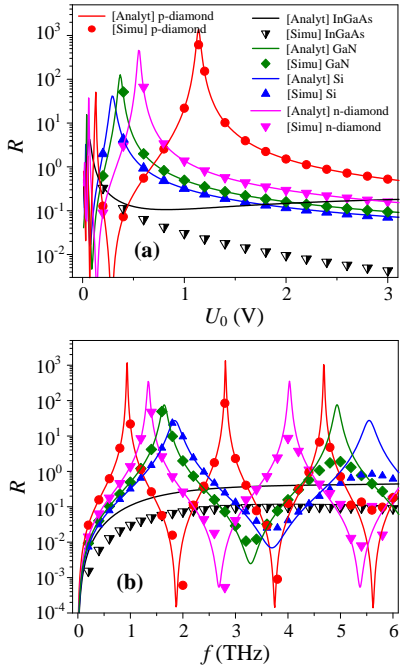

Fig. 7. Normalized response of different TeraFETs versus the gate bias $U_0$ (a) and frequency $f$ (b) under $T = 77$ K. The amplitude of incoming small signal is set to be 5 mV. The channel length $L = 130$ nm. The $f$ in (a) is fixed at 1 THz, and the $U_0$ in (b) is 1 V.

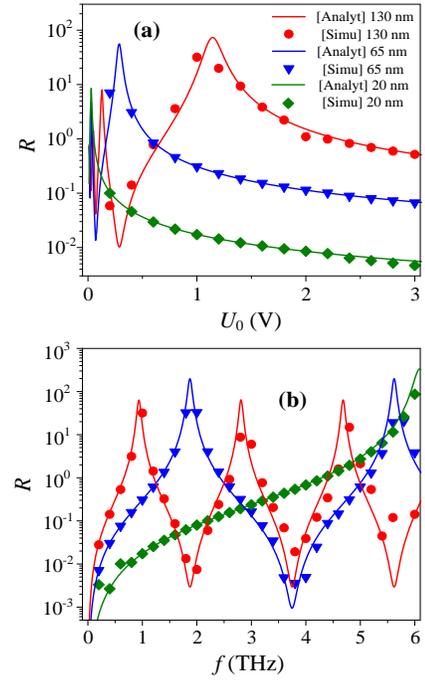

Fig. 8. Normalized response of the p-diamond TeraFET versus $U_0$ (a) and $f$ (b) under typical mobilities at 3 channel lengths. The $f$ in (a) is 1 THz, and the $U_0$ in (b) is 1 V. The amplitude of incoming small signal is 5 mV. The operating temperature is 300 K.

## C. Temperature dependence of TeraFET detectors

To evaluate the performance of plasma wave TeraFET detectors at different temperatures, we explore the response profiles of TeraFETs at the cryogenic temperature (77 K) and compare the results with those at room temperature (300 K). It should be pointed out that in the present hydrodynamic model, the energy equation only considers the heat transfer process inside the device [2, 3, 11]. In cryogenic temperature simulations, this could lead to a significant self-heating by the THz signal (i.e. the temperature level in the channel rises with time and saturates above the environment temperature). While in the real system, this heating effect may not be as significant since the heat exchange between the channel and ambient environment is always crucial. In our simulation, we assume temperature equilibrium for $T = 77$ K throughout the simulation period.

Fig. 6 plots the normalized DC response of p-diamond TeraFET versus $U_0$ and $f$ at 2 temperatures. As seen from Fig. 6(a), a significant elevation in peak response is achieved when the system temperature drops from 300 K to 77 K. This is mainly attributed to a substantial increase in carrier mobility as $T$ decreases from 300 K to 77 K. Fig. 6(b) illustrates the normalized DC response of p-diamond TeraFET as a function of frequency at 3 temperatures. Analogous to Fig. 6(a), the peak response increases as the temperature decreases from 300 K to 77 K. Those results suggest that the detector sensitivity could be even enhanced when the temperature drops.

To further demonstrate the effects of temperature variation on the detection characteristics, the DC response properties of TeraFETs at cryogenic temperature for other 4 materials are studied. Fig. 7 presents the gate bias and frequency dependence of response profiles for 5 types of TeraFETs under $T = 77$ K. Fig. 7(a) shows the gate bias dependence characteristics similar to those at $T = 300$ K, except that the response peaks are elevated due to the rise of mobility. The simulation data still conform to the analytical curves for p-diamond, n-diamond, Si and GaN TeraFETs, as the viscosity in those cases are still not too large to cause the distortion. Also, p-diamond TeraFET still has the highest peak response, and leads the response at $U_0 > 0.8$ V. Fig. 7(b) shows the frequency dependence of response profile at 77 K. From this figure we also observe a similar response characteristics and enhanced peak values compared to the 300 K case. Based on those results, it is clear that a cryogenic detection at $T = 77$ K does not alter the general properties of THz detectors, while the detection sensitivity is improved. Also, the p-diamond TeraFET still presents a superior performance at 77 K. Besides, a slight mismatch between simulation and analytical results shows up at 77 K for diamond, Si and GaN TeraFETs, since the simulation curve appears to be more or less flattened at higher frequencies, while the theory predicts an undamped oscillation. This phenomenon is cause by the high viscosity effect [3, 7], and is consistent with the previous observation by Rudin *et al.* [3].

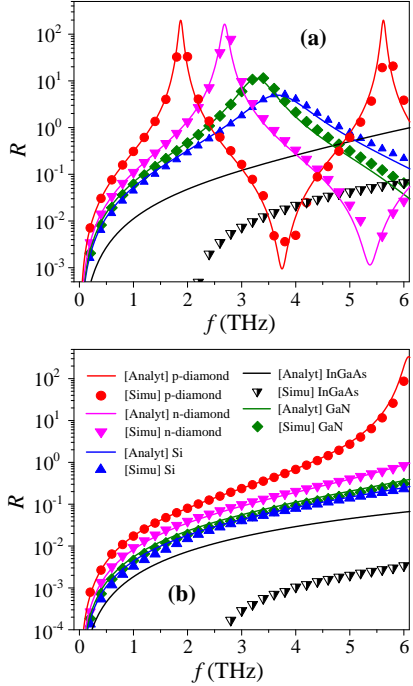

Fig. 9. Normalized response of the p-diamond TeraFET versus frequency under typical mobilities at channel lengths of 65 nm (a) and 20 nm (b). The DC gate bias $U_0$ is at 1 V, the amplitude of incoming small signal is 5 mV, and the operating temperature is 300 K.

*D. Channel length dependence of TeraFET detectors*

In this section, we discuss the effect of device channel length on the detection properties of TeraFETs. Fig. 8 presents the DC response of p-diamond TeraFETs versus $U_0$ and $f$ when the channel lengths are 130 nm, 65 nm and 20 nm at 300 K. As seen in Fig. 8(a), at $f$ = 1 THz, the peak response and general response level decrease as $L$ declines, indicating that a longer channel is preferred in this condition if a high responsivity is required. However, this rule may not hold under other frequencies. Fig. 8(b) illustrates the frequency spectrum of normalized response for p-diamond at 3 gated channel lengths. With the decrease of $L$, the maximum response rises. The rise of resonant response peak can be explained by the approximate expression of $f(\omega)$ in the vicinity of resonant peaks (i.e. $|\omega - n\omega_0| \ll \omega_0$, and $\omega\tau \gg 1$) [7]:

$$f(\omega) \approx 4(\frac{s\tau}{L})^2 \frac{1}{4(\omega-n\omega_0)^2\tau^2+1} \quad (9)$$

At resonant frequencies, $f(\omega) \sim 4(s\tau/L)^2$, which is inversely associated with $L$. Furthermore, the term $s\tau/L$ can also be used as the judgement for long/short devices. If $s\tau/L \ll 1$, the device can be regarded as a long device, and thus behaves like a broadband detector [7]. For $L$ equals to 130 nm, 65 nm and 20 nm, the values of $s\tau/L$ are 8.367, 16.734 and 54.387, respectively. Therefore, under the present mobility, the p-diamond TeraFET is always more qualified for being a resonant detector when $L \le 130$ nm. If we try to design a broadband detector, the channel length should be at least $s\tau$ = 1.09 μm. One can also acquire a broadband detector by reducing the mobility of the p-diamond sample. If the carrier mobility in p-diamond is dropped to 150 cm$^2 \cdot$V$^{-1}$s$^{-1}$ [4], the corresponding TeraFETs could serve as broadband detectors when $L$ is no less than 30.8 nm..

To compare the $L$ dependence of TeraFETs made with different materials, Fig. 9 plots the DC response of all 5 types of TeraFETs versus frequency under channel lengths of 65 nm and 20 nm at 300 K. As shown in Fig. 9(a), the relative response level of p-diamond TeraFET increases compared to the 130 nm case shown in Fig. 6(a). Furthermore, the frequency range at low frequency region within which the p-diamond device tops the response level expands. When $L$ is reduced to 20 nm, this "p-diamond dominant" region expands further to at least $f$ = 6 THz, as illustrated in Fig. 9(b). The extension of this frequency range is attributed to the increase of fundamental resonant frequency $f_0 = s/4L$ as $L$ drops. Moreover, it is also noticed that the magnitude of the response shown in Fig. 9(b) is negatively correlated to the effective massive of the detector material. In fact, this rule is quite universal and holds in all the sub-THz regions shown in other figures (e.g. Fig. 5(a) and Fig. 7(b)). To explore the mechanism behind this rule, we consider the analytical response equation in this region. We first consider the low frequency region in which $\omega\tau \ll 1$ and $k_0' \approx k_0'' \approx (1/s)\cdot(\omega/2\tau)^{0.5} = k_0$, where $k_0'$ and $k_0''$ are the real and imaginary part of wave vector $k$ for plasma wave propagation [3, 7]. In this region, we always have $k_0 L \ll 1$, thus $\sin^2(k_0 L) \approx 0$ and $\cos^2(k_0 L) \approx 1$, then the expression of $f(\omega)$ becomes [7]:

$$f(\omega) = \frac{\sinh^2(k_0 L) - \sin^2(k_0 L)}{\sinh^2(k_0 L) + \cos^2(k_0 L)} \approx \frac{\sinh^2(k_0 L)}{\sinh^2(k_0 L)+1} \quad (10)$$
$$= \frac{1}{1+1/\sinh^2(k_0 L)}$$

Note that $k_0 \sim 1/s$ thus $k_0 \sim m^{0.5}$. With a larger effective mass, $\sin^2(k_0 L)$ increases, leading to the increase of $f(\omega)$ and therefore the magnitude of DC response.

Now we further consider the region where $\omega\tau > 1$ and $\omega < \omega_0$. In this region, $k_0' \approx \omega /s$, $k_0'' \approx 1/2s\tau$. For the case of $L$ = 20 nm, we have $k_0'' L = L/2s\tau \ll 1$, thus $f(\omega)$ can be approximately expressed by [7]:

$$f(\omega) = \frac{3\sinh^2(k_0'' L) + \sin^2(k_0' L)}{\sinh^2(k_0'' L) + \cos^2(k_0' L)} \quad (11)$$
$$\approx \frac{\sin^2(k_0' L)}{\cos^2(k_0' L)} = \tan^2(k_0' L)$$

Since $k_0' \sim 1/s \sim m^{0.5}$, $f(\omega) = \tan^2(k_0'' L)$ will be positively associated with the carrier effective mass. Therefore, in both high frequency and low frequency regions, a higher carrier effective mass yields a higher DC response if the channel is short enough. Those discoveries suggest that p-diamond TeraFETs may be more advantageous as a high sensitivity detector under relatively small feature sizes.

## IV.  CONCLUSION

The hydrodynamic simulations show that the p-diamond TeraFETs can operate at the resonant mode within sub-terahertz frequency range, and have a relatively low critical resonant mobility, thus very suitable to be used as resonant detectors. Under the high gate bias and low frequency (sub-THz), the high mobility p-diamond TeraFETs have a higher DC voltage response compared to other TeraFETs. As the temperature drops from 300 K to 77 K, the general detection characteristics do not alter, while the amplitude of the response enhances, suggesting the feasibility of THz detection using plasmonic detectors at 77 K. At both room temperature and 77 K, p-diamond presents the largest peak response and leads all TeraFETs in DC response in a large dynamic range.

As the channel length reduces to 20 nm, the p-diamond TeraFET exhibits the highest DC response among all types of TeraFETs in a large frequency range. This indicates that the relative performance of p-diamond TeraFETs improves as the feature size scales down.